**Comment on "A compilation and bioenergetic evaluation of syntrophic microbial growth yields in anaerobic digestion" by Patón, M. and Rodríguez, J. [Water Research 162 (2019), 516–517]**

Helena Junicke[*], PhD, Assistant Professor

Department of Chemical and Biochemical Engineering, Technical University of Denmark, Søltofts Plads, Building 229, 2800 Kgs. Lyngby, Denmark

[*]Corresponding author. Email: heljun@kt.dtu.dk

**Abstract**

Recent efforts have focused on providing a systematic analysis of syntrophic microbial growth yields. These biokinetic parameters are key to developing an accurate mathematical description of the anaerobic digestion process. The agreement between experimentally determined growth yields and those obtained from bioenergetic estimations is therefore of great interest. Considering five important syntrophic groups, including acetoclastic and hydrogenotrophic methanogens, as well as propionate, butyrate and lactate oxidizers, previous findings suggested that measured and estimated growth yields were consistent only for acetoclastic methanogens. A re-analysis revealed that data are also consistent for lactate oxidizers and hydrogenotrophic methanogens, whereas the limited data available for propionate and butyrate oxidizers are unsupportive of firm conclusions. These results highlight pertinent challenges in the analysis of microbial syntrophy and encourage more accurate measurements of syntrophic microbial growth yields in the future.

## 1. Introduction

Water Research has recently published an article entitled "A compilation and bioenergetic evaluation of syntrophic microbial growth yields in anaerobic digestion" (Patón and Rodríguez, 2019). The article compares experimentally observed growth yields to those estimated via the Gibbs energy dissipation method (Heijnen and Dijken, 1992). Focus is given to five important syntrophic groups, namely acetoclastic and hydrogenotrophic methanogens, as well as propionate, butyrate and lactate oxidizers. Patón and Rodríguez (hereafter: PR) conclude that "[o]nly for acetoclastic methanogens the bioenergetic estimation of microbial growth yields […] appeared consistent with experimental observations". A re-analysis of the original data reveals that these findings have to be revised.

## 2. Growth yields of lactate oxidizers appear consistent

PR found measured and estimated yields of lactate oxidizers inconsistent, but these conclusions change when considering the following corrections: (1) The growth yield reported by Junicke et al. (2016) was wrongly taken as 0.143 gCOD-X/gCOD-Lac. The correct value is 0.048 gCOD-X/gCOD-Lac, which deviates from the estimated yield by merely 2%; (2) The growth yield reported by Traore, Fardeau, et al. (1983) is 0.077 gCOD-X/gCOD-Lac, about half as large as initially presented. The initially larger value of 0.147 gCOD-X/gCOD-Lac relates to Traore, Gaudin, et al. (1983); (3) The determination of individual growth yields in mixed microbial cultures requires special care. If such yield is based on a lumped biomass increase, it assumes that all biomass formed is due to a single species. This is not correct. A so-derived growth yield may be substantially overestimated. The yield reported by Wallrabenstein et al. (1995) is a case in point and should be excluded.

Considering these amendments, experimental and estimated yields appear identical within one standard deviation in three out of four cases (see Fig. 1a), demonstrating the exact opposite of PR's conclusion.

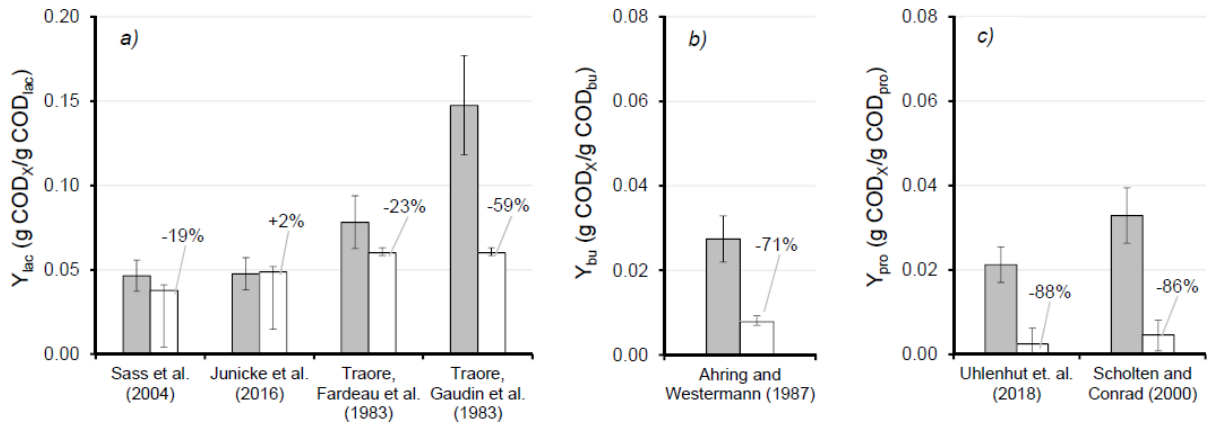

**Fig. 1:** Microbial growth yields for lactate (a), butyrate (b), and propionate oxidizers (c). Experimentally determined values (gray) and bioenergetically estimated values (white).

### 3. Limited data prevent conclusions on propionate and butyrate oxidizers

The microbial growth yields of propionate and butyrate oxidizers deserve further attention. (1) The growth yield on propionate reported by Wallrabenstein et al. (1995) is a lumped yield and should be excluded given its significant risk of overestimation; (2) The yield on propionate reported by Scholten and Conrad (2000) is doubtful due to conflicting unit conversions. (3) The individual growth yields on butyrate and propionate referring to Lawrence and McCarty (1969) have been presented as if they were experimentally determined yields, but in fact, neither of them was directly measured in that study. PR must have estimated the individual growth yields based on the reported total biomass yield. Consequently, they are estimated yields, not measured ones, which precludes them as a valid reference for just another estimation method. (4) Another critical aspect is the use of generic conversion factors. The growth yield obtained from Beaty and McInerney (1989) is one example. The butyrate-consuming species was grown in a co-culture and its growth quantified by manual cell counting. However, the study provides no correlation between cell numbers and biomass, thus giving insufficient data to determine the growth yield. To obtain the individual yield, PR must have assumed a conversion factor, but this approach is highly critical. It is a sensitive parameter that depends on the species, cell size and geometry, as well as the physiological status of the cell. To enable a better approximation of uncertainty, it is instructive to consider an idealized spherical cell of homogeneous mass density and 2.0 µm diameter. The cell has a volume of 4.18 µm³. If the cell diameter was off by only 10%, i.e. 2.2 µm, the volume becomes 5.57 µm³ – an error in cell mass of already 33%. However, size alone is not a reliable predictor for cell mass. The cell dry mass of 2-µm-long *E. coli* cells is typically estimated as 0.3 pg/cell. Comparing this to 7-µm-long *M. hungatei* cells, one would expect a considerably larger cell dry mass, but the reported value is much lower, approximately 0.1 pg/cell (Scholten and Conrad, 2000). Therefore, even slight differences in cell size or composition can lead to substantial errors, which means the so-calculated growth yield should be excluded.

These revisions reduce the amount of reliable data from three samples to just one (butyrate oxidizers), and from four to only two samples (propionate oxidizers), while still counting Scholten and Conrad (2000) as a valid source (see Fig. 1b and 1c). Therefore, to allow firm conclusions on the growth yields of propionate and butyrate oxidizers, more data are needed, together with a detailed elaboration of measurement uncertainty.

**4. Yield differences for hydrogenotrophic methanogens are statistically insignificant**

A paired-samples t-test was conducted to evaluate whether a significant difference exists between measured and estimated growth yields of hydrogenotrophic methanogens (95% confidence level). The observed differences (Mean = 0.0072 gCOD-X/gCOD-$H_2$, Std.Dev = 0.0179 gCOD-X/gCOD-$H_2$, N = 10) are statistically insignificant (t = 1.26, df = 9, two-tail p = 0.239), thus rejecting PR's conclusion and confirming instead that estimated growth yields are consistent with experimental observations.

**5. Measurement uncertainty requires detailed attention**

It is worth noting that the yield comparison contains no indication of measurement uncertainty for the reported experimental growth yields. Following scientific standard approaches (e.g.

ISO GUM), a measurement result without quantitative statement of its uncertainty does not qualify for comparisons. PR have underlined that the quantification of growth yields is a delicate task, the methods employed often error-prone and subject to individual inaccuracies. A proper evaluation of measurement uncertainty appears indispensable as otherwise the yield comparison lacks a solid foundation. In the presented re-analysis, a standard deviation of 20% was assumed for all experimental yields. This is in line with other works, e.g. Traore, Fardeau, et al. (1983), while noting that actual errors might be considerably larger for the individual studies. For example, manual cell counting entails standard deviations of up to 40% (Scholten and Conrad, 2000). A more rigorous quantification approach is presented by Junicke et al. (2014).

## 6. Conclusions

The recent article by Patón and Rodríguez (2019) contributes to a growing body of knowledge on anaerobic digestion. It is thus relevant to scholars aiming to uncover the fundamental mechanisms of syntrophy, and industrial operators active in wastewater treatment or biogas production. A re-analysis of the original data revealed several critical aspects that necessitate the following amendments:

- Measured growth yields and bioenergetic estimations are consistent for syntrophic lactate oxidizers as well as for hydrogenotrophic and acetoclastic methanogens
- Consistency can neither be rejected nor confirmed for propionate and butyrate oxidizers due to a lack of reliable experimental evidence
- Consideration of measurement uncertainties is essential to ensure the significance of syntrophic growth yield comparisons

With this comment it is aimed to highlight pertinent challenges related to individual biomass quantification, hoping that it will encourage more accurate measurements of syntrophic microbial growth yields in the future.

**Acknowledgements**

The author acknowledges the support obtained from the European Union's Horizon 2020 research and innovation programme under the Marie Sklodowska-Curie grant agreement number 713683 (COFUNDfellowsDTU); the Danish Council for Independent Research (grant number 7017-00175A); and the Novo Nordisk Fonden in the frame of the Fermentation-Based Biomanufacturing initiative. The author would like to thank Mauricio Patón and Jorge Rodríguez for the valuable discussion.